\documentclass[12pt]{iopart}

\usepackage{graphicx}
\usepackage{bm}
\usepackage{amssymb}
\usepackage{epsfig}
\usepackage[usenames]{color}
\begin{document}

\title{Double-gated graphene-based devices}

\author{S Russo$^{1,2}$, M F Craciun$^1$, M Yamamoto$^1$, S Tarucha$^{1,3}$ and A F Morpurgo$^{4}$}
\address{$^1$ Department of Applied Physics, The University of Tokyo,
Tokyo 113-8656, Japan}
\address{$^2$ Kavli Institute of Nanoscience, Delft
University of Technology, Lorentzweg 1, 2628 CJ Delft, The
Netherlands}
\address{$^3$ Quantum Spin Information Project,
ICORP, Japan Science and Technology Agency, Atsugi-shi, 243-0198,
Japan}
\address{$^4$ DPMC and GAP, University of Geneva, quai
Ernest-Ansermet 24, CH-1211 Geneva 4, Switzerland}
\ead{s.russo@tnw.tudelft.nl}

\begin{abstract}

We discuss transport through double gated single and few layer
graphene devices. This kind of device configuration has been used to
investigate the modulation of the energy band structure through the
application of an external perpendicular electric field, a unique
property of few layer graphene systems. Here we discuss
technological details that are important for the fabrication of top
gated structures, based on electron-gun evaporation of SiO$_2$. We
perform a statistical study that demonstrates how --contrary to
expectations-- the breakdown field of electron-gun evaporated thin
SiO$_2$ films is comparable to that of thermally grown oxide layers.
We find that a high breakdown field can be achieved in evaporated
SiO$_2$ only if the oxide deposition is directly followed by the
metallization of the top electrodes, without exposure to air of the
SiO$_2$ layer.

\end{abstract}

\maketitle

\section{Introduction}

The ability to isolate and embed  single- and multi-layer graphene
in double gated structures is paving the way to reveal unique
electronic properties of these systems
\cite{graphene1,graphene2,graphene3,Kleinparadox,graphene4,graphene5,graphene6,bilayer1,bilayer2,bilayer3,bilayer4,trilayer1,trilayer2}.
The ability to change the voltages applied to a nano-fabricated top
gate and to the back-gate offers the possibility to gain local
control of the charge density and of imposing locally a
perpendicular electric field. This device configuration was used
recently to show how the band structure of graphene-based materials
can be tuned continuously \cite{bilayer1,trilayer1}. In particular,
bilayer graphene exhibits an electric field induced insulating state
due to the opening of a gap between valence and conduction band
\cite{bilayer1}, and in trilayers, which are semi-metals, the band-overlap can be increased
substantially \cite{trilayer1}. Cleverly designed top gates on a
graphene single layer have also been used successfully for
engineering p-n junctions
\cite{graphene1,graphene2,graphene3,graphene4,graphene5,graphene6},
necessary for the investigation of Klein tunneling
\cite{graphene1,graphene3,Kleinparadox}, and to attempt the
fabrication of controlled quantum dots \cite{QD1,QD2}. Key to the
fabrication of top gated structures is the ability to deposit good
quality thin gate oxides, with high breakdown field and low
leakage current.\\
Here, after reviewing the relevance of double gated devices on the
electric field modulation of the band structure of double and triple
layer graphene, we discuss in some details technological aspects
related to the properties of the SiO$_2$ layers used as gate
insulators. In particular, we discuss how we can routinely achieve
high breakdown fields in electron-gun evaporated thin SiO$_2$ films
(15 nm), comparable to the breakdown fields of thermally grown
SiO$_2$, which is surprising given that SiO$_2$ deposited by
evaporation was long believed to be a poor quality insulator. To
unveil the reasons behind the good insulating quality of our
evaporated SiO$_2$ films, we conducted a statistical study of
leakage current and breakdown voltage in capacitors, where two
metallic electrodes are separated by a SiO$_2$ layer fabricated in
different ways. We demonstrate that if SiO$_2$ and top gate metal
electrodes are deposited subsequently without exposing the SiO$_2$
to air, the electrical performance of electron-gun evaporated
SiO$_2$ is comparable to that of thermally grown SiO$_2$. In
contrast, exposure to air of the SiO$_2$ layer before deposition of
the counter-electrode leads to much worse insulating
characteristics. Our findings indicate that extrinsic degradation
--probably due to the absorption of humidity- has limited in the
past the insulating quality of electron-gun evaporated SiO$_2$.

\section{Device and fabrication}

\begin{figure}[h]
\begin{center}
\includegraphics[width=\linewidth]{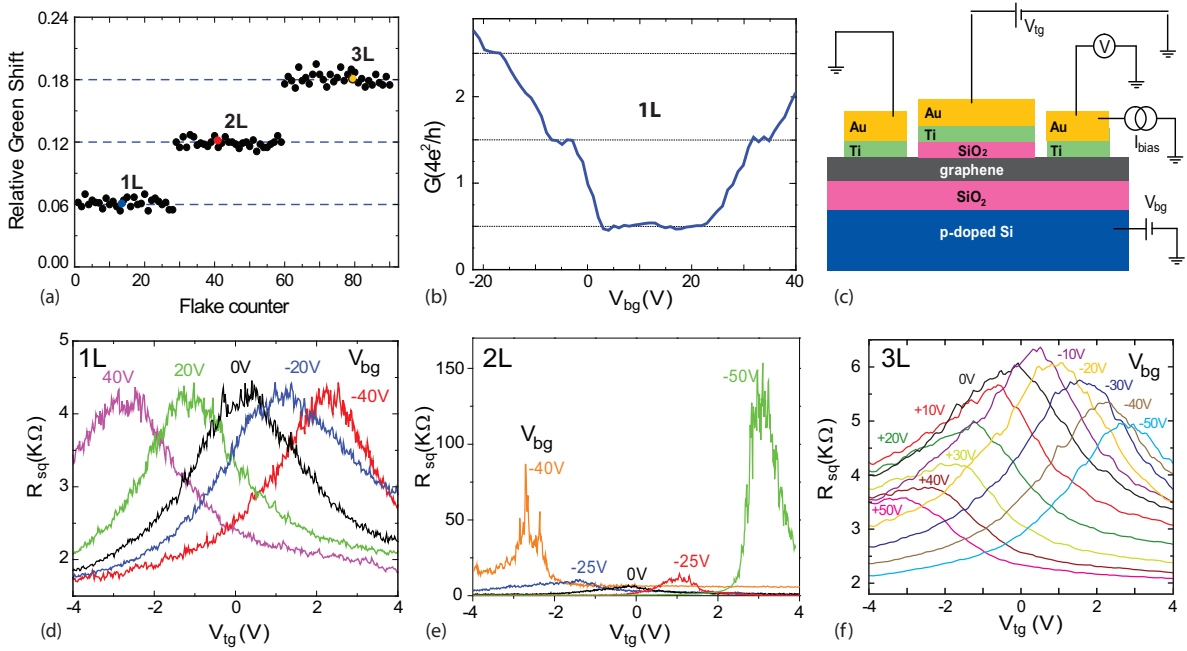}
\caption{Panel (a) shows a plot of $\textit{Relative Green Shift}$
(as defined in the main text) for 90 few layer graphene areas,
showing clear plateaus for graphene layer of different thickness.
Panel (b) shows the conductance of a device (indicated by the blue
dot in the plot shown in panel (a)) measured in the presence of high
magnetic field (B=8T); plateaus are observed at conductance values
characteristic for Dirac electrons, indicating that flakes with
$\textit{Relative Green Shift}=0.06$ are indeed single layers. Panel
(c) shows a schematic representation of double-gated graphene
devices, together with the measurement configuration used to study
transport as a function of voltages applied to the top and back
gates. Panels (d), (e) and (f) show plots of the square resistance
in single (d), double (e) and triple (f) layers respectively,
measured while sweeping the top gate, for different fixed
voltages applied on the back gate (for single layer $T=300mK$, $W=2
\mu m$, $L=1.4 \mu m$, $\mu=3500cm^2/Vs$, double layer $T=300mK$,
$W=1.7\mu m$, $L=1\mu m$, $\mu=900cm^2/Vs$ and for triple layer
$T=1.5K$, $W=1 \mu m$, $L= 1.4 \mu m$, $\mu=800 cm^2/Vs$).}
\label{Russo_fig1}
\end{center}
\end{figure}

Single and few layer graphene flakes used for the device
fabrication were obtained by micro-mechanical cleavage of natural
graphite crystals, and by their subsequent transfer onto a highly
doped Silicon substrate (acting as a gate) covered by a $285$ nm
thick thermally grown SiO$_2$ layer. The thickness of the graphene
layers can be simply identified by analyzing  the shift in
intensity in the RGB green channel relative to the substrate (i.e.
\textit{Relative Green Shift})
\cite{bilayer1,trilayer1,graphenevisibility1,graphenevisibility2,graphenevisibility3}.
A plot of the relative green shift, as extracted from optical
microscope images of various samples taken with a digital camera,
exhibits plateaus corresponding to the discrete thickness values -see Fig. \ref{Russo_fig1}a. Subsequent
transport measurements (quantum Hall effect, resistance dependence
on a perpendicular electric field, etc.) confirm the validity of
this optical method (Fig. \ref{Russo_fig1}b).\\
The fabrication of nanostructures is accomplished by conventional
electron-beam lithography. Metallic contacts and top gates were
deposited by electron-gun evaporation, respectively of Ti/Au
($10/25$nm thick with a back ground pressure of $9*10^{-7}$torr) for
contacts and SiO$_{2}$/Ti/Au ($15/10/25$nm thick with a back ground
pressure of $3*10^{-7}$ torr) for top gates, followed by lift-off.
We took special care to fabricate all the ohmic contacts within 60
nm from the edges of the top gated areas, so that two probe
resistance measurements are dominated by the resistance of the
double gated region\cite{contact res}. All transport measurements in
double gated devices (see Fig. \ref{Russo_fig1}c) were made using a
lock-in technique (excitation frequency: $19.3$Hz), in the linear
transport
regime, at temperatures ranging from 300 mK up to 150K.\\
To understand why, contrary to expectations, we manage to achieve
high breakdown field in thin, electron-beam evaporated SiO$_2$
films, we conducted a macroscopic study of the breakdown characteristics on two types of
capacitor test structures. The first -which we refer to as
\textit{type A}- is characterized by subsequent evaporation of
SiO$_2$/Ti/Au without breaking the vacuum in between the deposition
of the different materials. For the second -$\textit{type B}$
SiO$_2$- we exposed the device to ambient for one hour after the
SiO$_2$ deposition, before evaporating the Ti/Au counter electrode.
The breakdown test measurements were made with a Keithley-2400
source-meter on more than 130 different capacitors (with three
different surface areas: $125 \times 115 \mu m^2$, $175 \times 150
\mu m^2$ and $215 \times 195 \mu m^2$).\\

\section{Transport experiments in double gated few layer graphene devices}

\begin{figure}[h]
\begin{center}
\includegraphics[width=\linewidth]{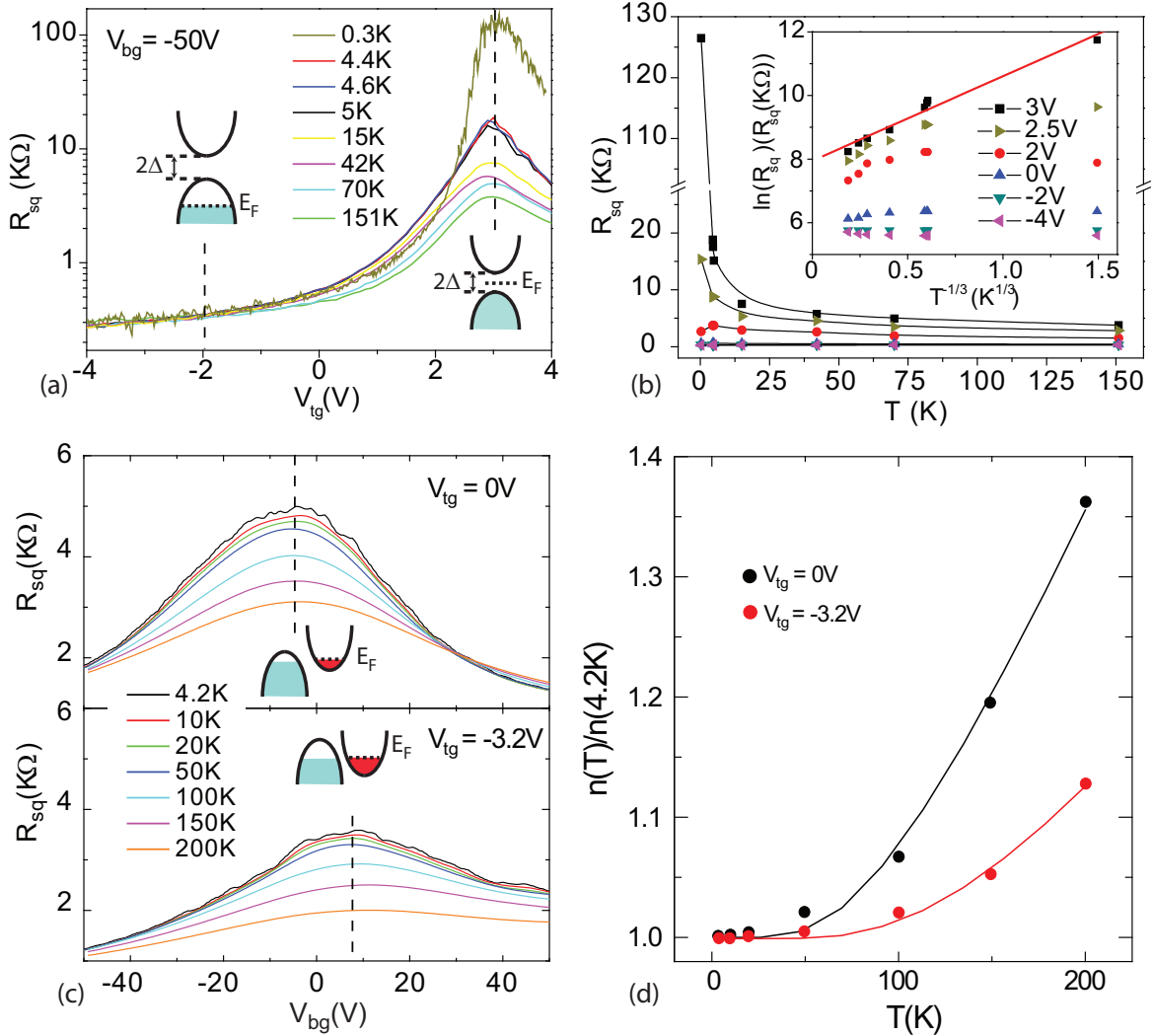}
\caption{Panels (a) and (c) show plots of the square resistance at
different temperatures for the bilayer and trilayer devices whose
data are shown in panel (e) and (f) of Fig.\ref{Russo_fig1}. Panel
(b) shows a plot of $R_{sq}$ $\textit{vs.}$ $T$ and $ln(R_{sq})$
$\textit{vs.}$ $T^{1/3}$ extracted from the measurements shown in
panel (a), at $V_{bg}=-50V$ and different $V_{tg}$ values as
indicated in the legend. The external electric fields applied on the
bilayer for each different $V_{tg}$ are respectively -0.177 V/nm
($\blacksquare$), -0.175 V/nm
($\textcolor{LimeGreen}{\blacktriangleright}$), -0.173V/nm
($\textcolor{red}{\bullet}$), -0.167 V/nm
($\textcolor{MidnightBlue}{\blacktriangle}$), -0.16 V/nm
($\textcolor{Cyan}\blacktriangledown$), -0.153 V/nm
($\textcolor{Mulberry}{\blacktriangleleft}$). Panel (d) shows the
normalized charge density as a function of temperature, extracted
from the trilayer measurements of panel (c). The continuous line is
a fit based on a two band model with finite overlap (see main
text).} \label{Russo_fig2}
\end{center}
\end{figure}

The measurement configuration used for double gated devices is shown
in Fig. \ref{Russo_fig1}c. A finite voltage applied to either one of
the gates (back or top gate) changes the position of the Fermi level
in the corresponding gated region of the graphene layer, by an
amount corresponding to the induced charge density. In addition, by
biasing the two gate electrodes with opposite polarity, a large
external electric field applied perpendicular to the  layer is
generated, which is equal to $E_{ex}=(V_{bg}-V_{tg})/d_{tot}$
($d_{tot}=15+285 nm$ is the total SiO$_2$ thickness). In this device
configuration, we can monitor the evolution of the in plane
transport properties for each
few layer graphene device as a function of $E_{ex}$.\\
Fig. \ref{Russo_fig1}d, e and f show the typical behavior of the
square resistance measured respectively in double gated single (d),
double (e) and triple (f) layer graphene, when sweeping the top gate,
while keeping the back gate at a fixed potential. It is apparent that
the overall electric field dependence of $R_{sq}$ is markedly distinct
for graphene layers of different thickness. In all cases, the resistance exhibits a maximum
($R_{sq}^{max}$) whose value and position in gate voltage depend on
the voltage applied to the gate on which a fixed potential is
applied during the measurement. At $E_{ex}=0 $ V/m we find that for
single and bilayer graphene $R_{sq}^{max} \sim 6 K\Omega$, close to
a conductance per square of  $4e^2/h$, as expected, indicating that
the fabrication of top gate structures does not damage significantly
the material (for trilayers the square resistance is somewhat lower,
owing to the presence of an overlap between valence and conduction
band). Increasing the external electric field induces a well defined
-and different- response for the square resistance of layers of different
thickness. Respectively, in a single layer $R_{sq}^{max}$
is not affected by a finite $E_{ex}$; in bilayers at low temperature
$R_{sq}^{max}$ increases from $6 K\Omega$ to very large values ($>
100$ K$\Omega$); in trilayers $R_{sq}^{max}$ decreases with increasing
$E_{ex}$. These experimental findings, confirmed in a number of
different samples (3 single layers, over 10 double and 10 triple
layers) provide a clear indication that each few layer graphene is a
unique material system, with distinct electronic properties.\\
Transport measurements over a wide range of temperatures (from
$300mK$ up to $150K$) underline the unique electronic properties of
these few layer graphene devices. In bilayer graphene the larger
$E_{ex}$, the more pronounced is the temperature dependence of
$R_{sq}^{max}$, see Fig. \ref{Russo_fig2}a and b. At $E_{ex}=0$ V/m,
$R_{sq}^{max}$ is only weakly temperature dependent (as it is
typical of zero-gap semiconductors), and at $E_{ex}\neq 0$ V/m the
observed behavior is the one typical of an insulating state. On the
contrary, trilayer graphene devices display a decrease of
$R_{sq}^{max}$ when lowering the temperature, stemming for the
semimetallic nature of the constituent material.\\
At a more quantitative level we find that $R_{sq}^{max}$ in bilayer
graphene is well described by $\propto exp(T_{0}/T)^{1/3}$, with
$T_{0} \approx 20$ K at the maximum applied external electric field
(see inset in Fig. \ref{Russo_fig2}b). This temperature dependence
is indicative of variable-range hopping in a two dimensional
material where an energy gap has opened, and where disorder causes
the presence of sub-gap states ($T_{0}$ is related to this subgap density of states\cite{bilayer1})
-making it difficult to estimate $\Delta$ from transport experiments.\\
Indeed, in a disorder free bilayer graphene, at
$E_{ex}\neq 0$ a gap ($\Delta$) opens in the band structure and
the density of states in the gapped region is zero. In this ideal
case, when the Fermi level is in the middle of the gapped region the value of $R_{sq}^{max}$
at a finite temperature is
entirely determined by thermally activated charge carriers ($R_{sq}^{max} \propto \exp(\Delta/{2 k_B T})$).
Therefore, $\Delta$ can be accurately determined from a plot of
$ln(R_{sq}^{max})$ $\textit{versus}$ $1/T$. On the other hand, in real devices the presence of
disorder creates a finite density of states in the the band gap of bilayer
graphene. Now charge carriers can conduct via variable-range hopping at $R_{sq}^{max}$. In this case
$R_{sq}^{max}$ is a
function of the density of states at the Fermi level and not any more simply a function of $\Delta$.\\
The fact that a gap opens in the band structure of bilayer graphene at
finite $E_{ex}$ is evident from the
temperature dependence of $R_{sq}$ as a function of charge density.
In particular, when the Fermi level lays deep into the conduction and/or valence
band, a temperature independent $R_{sq}$ is expected.
However, when the Fermi level is shifted across the energy gap region $R_{sq}$
should display an insulating temperature dependent behavior. Experimentally
this is achieved by measuring $R_{sq}$ at a fixed value of
either of the gates (e.g. $V_{bg}$) and for different voltage
applied on the other gate (e.g. $V_{tg}$) (see Fig \ref{Russo_fig2}a and b). The cross over from band transport
to variable hopping range in the gap occurs  at the edge of the valence and conduction band (see Fig. \ref{Russo_fig2}b).\\
Similar previous transport experiments in double gated bilayer \cite{bilayer1}
reported an energy scale of 1-10mV associated with the insulating state induced by $E_{ex}\neq 0$.
This energy scale seen in transport
is much smaller than the energy gap recently probed in optical spectroscopy experiments
($\Delta \sim 200mV $ at $E_{ex}= 2V/nm$ \cite{Crommie}). Possibly the finite sub-gap density of states
induced by the disorder is at the origin
of the small energy scale measured in transport experiments,
however the specific mechanism responsible for these experimental observations remains an open question.\\
The temperature dependence of the resistance in trilayer graphene is
opposite to the one observed in bilayers, and it reveals that this
material system is a semimetal with a finite overlap ($\delta
\varepsilon$) between conduction and valence band. This band overlap
can be estimated within a two band model
\cite{trilayer1,2bandmodel,FLG}, where the number of thermally
excited carriers increases with temperature according to $n(T)=(16
\pi m^*/h^2c)k_B T ln(1+e^{\delta \varepsilon /2 K_B T})$ ($m^*$
effective mass and $c$ equal to twice the layer spacing).
Measurements at finite $E_{ex}$ show that $\delta \varepsilon$
decreases when increasing external electric field ($\delta
\varepsilon$ goes from 32 meV to
52 meV in the measurements of Fig. \ref{Russo_fig2}d).\\
These experiments demonstrate that a perpendicular electric field
applied on few layer graphene is a valuable tool to change the band
structure of these materials. Double gated structures lead to the
discovery of the only known electric field tunable insulator, i.e.
bilayer graphene, and of the only known electric field tunable
semimetal, i.e. trilayer graphene.\\

\section{Evaporated Silicon oxide as top gate dielectric}

\begin{figure}[h]
\begin{center}
\includegraphics[width=\linewidth]{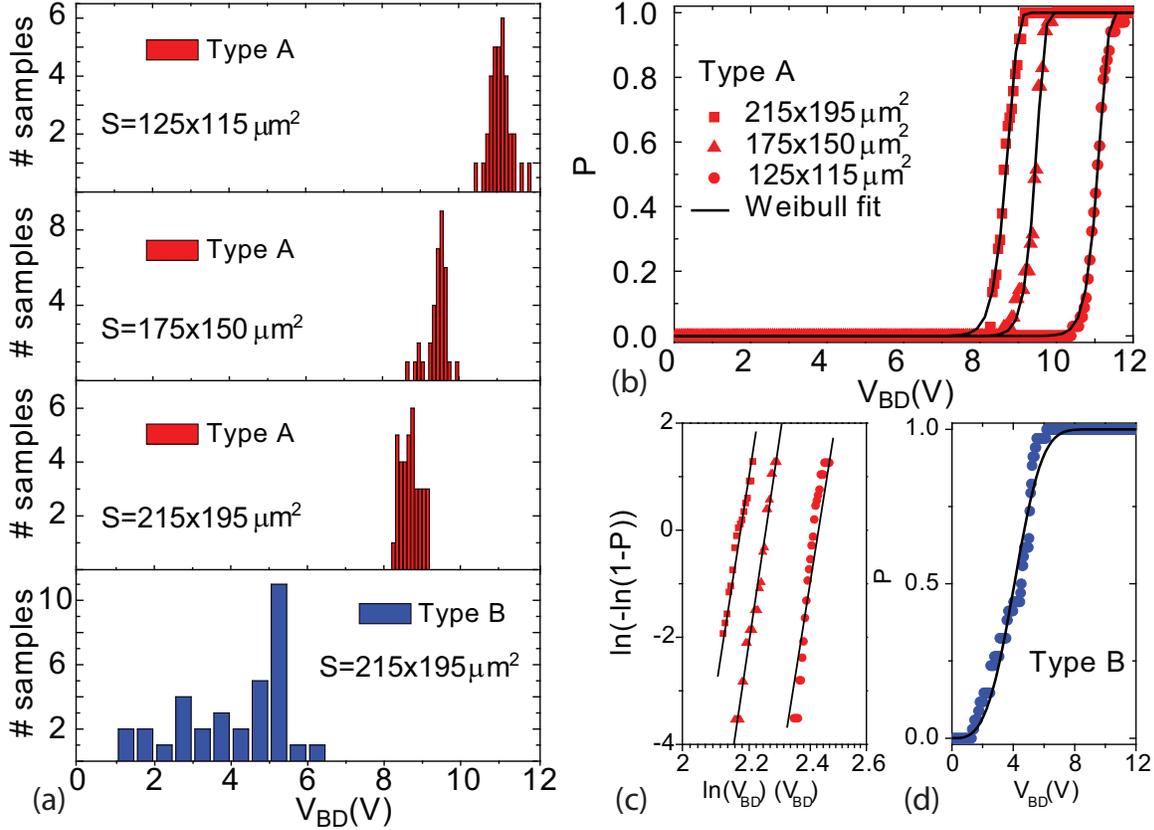}
\caption{Panel (a) shows histogram plots of the breakdown field for
different populations of test devices as specified in the legends.
The panels in (b) and (c) show the cumulative probability for
$\textit{type A}$ devices for different surface areas (dots are
experimental data, and the continuous line is a fit to the Weibull
distribution). The plot in d) is a fit to the cumulative probability
for $\textit{type B}$ devices ($S=215\times195 \mu m^2$) to the
Weibull distribution.} \label{Russo_fig4}
\end{center}
\end{figure}

The opening of a sizeable band gap in bilayer graphene, and large changes in the band
overlap of graphene trilayers require the application of large external electric fields
to these material systems. It is the breakdown field of the gate-dielectric that imposes a
limit to the maximum value of $E_{ex}$ experimentally accessible in double gated structures. To
optimize this aspect of the devices, we conducted a systematic study
of the breakdown electric field of SiO$_2$ gate oxide for devices
with different areas, fabricated under different conditions. Here
we discuss the details of this investigation. From our statistical analysis
we conclude that what is crucial is not the SiO$_2$ deposition
method, but the details in the metallization of top gate electrodes
afterwards, which affect the final quality of the oxide gate dielectric.\\
\begin{figure}[h]
\begin{center}
\includegraphics[width=\linewidth]{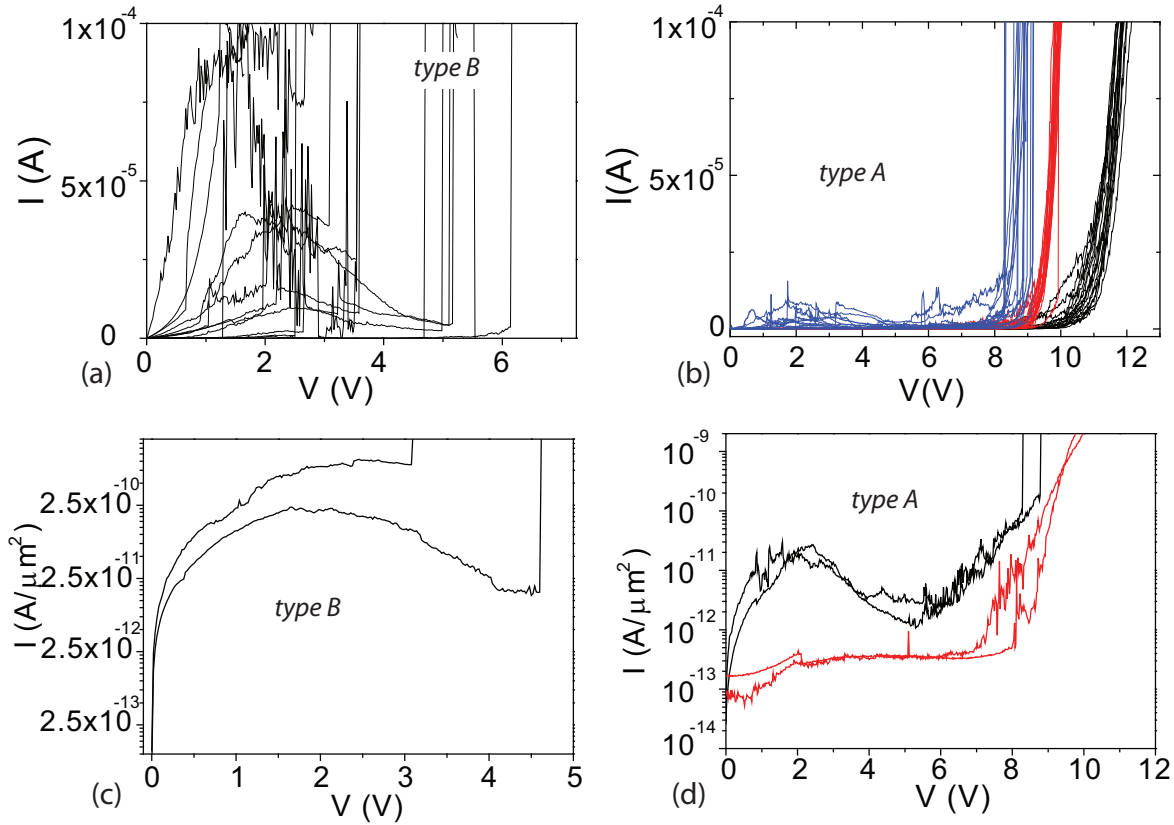}
\caption{Panels (a) and (b) show $I-V$ characteristics of
respectively $\textit{type B}$ devices with $S=215\times 195 \mu
m^2$ (a), and $\textit{type A}$ devices with $S=215\times 195 \mu
m^2$ (blue curves), $S=175\times 150 \mu m^2$ (red curves) and
$S=125\times 115 \mu m^2$ (black curves) (b). Panels (c) and (d)
show the $I-V$ curves in logarithmic scale for the two
$\textit{types}$ of devices and areas $S=215\times 195 \mu m^2$
(black curves in (c) (d)) and $S=175\times 150 \mu m^2$ (red curves
in (d)).} \label{Russo_fig3}
\end{center}
\end{figure}
We compare capacitor devices fabricated following two different
procedures for one of the oxide dielectric/metal electrode
interfaces. In particular, the devices were
fabricated on a Si/SiO$_2$ substrate (identical to the one used for
the graphene devices previously described) on which we deposit a
Ti/Au (5/20 nm) film -common electrode for the
capacitors. In devices of $\textit{type A}$, the SiO$_2$ deposition
and top electrode (Ti/Au 5/20 nm) metallization processes were
carried without breaking the vacuum. On the contrary in devices of
$\textit{type B}$, the SiO$_2$ gate dielectric was exposed to air
for one hour prior to the deposition of the top electrode metals. The SiO$_2$ deposition
was carried out typically at $3*10^{-7}$ torr back ground pressures. We did not observe
a dependence of the insultating properties of the SiO$_2$ dielectric,
breakdown field and leakage current over the background pressure range from $1*10^{-7}$ to $5*10^{-7}$torr.\\
Fig. \ref{Russo_fig3}a and b show various $I-V$ traces, measured in
ambient condition, for \textit{type A} and \textit{type B} devices.
A first clear difference between the two types of devices is the
magnitude of the leakage current, visible by plotting the $I-V$
curves both in linear and logarithmic scale, see Fig.
\ref{Russo_fig3}. For a surface area of $215 \times 195 \mu m^2$, we
find $I_{leakage} \simeq 2.5*10^{-10} A / \mu m^2$ for the best
$\textit{type B}$ devices which is one order of magnitude larger
than that measured in the worst $\textit{type A}$ devices (the
difference for typical devices are much larger than one order of
magnitude). This extremely different level of leakage current
already indicates that the exposure of SiO$_2$ to air previous to
the deposition of top metals has a large negative influence on the
insulating performance
of the oxide.\\
The $I-V$ characteristics further show that for a fixed surface area
($215 \times 195 \mu m^2$), the breakdown voltage $V_{BD}$ for
\textit{type A} is typically in the range $8V<V_{BD}<9V$ whereas
\textit{type B} devices breakdown anywhere in the range
$0V<V_{BD}<6V$. The differences in the failure of device types are
best summarized in the histogram plots of $V_{BD}$ -see Fig.
\ref{Russo_fig4}a. For $\textit{type B}$ we find a large spread
in the distribution of $V_{BD}$, in contrast to the narrow
distribution characteristic of $\textit{type A}$ devices.
Furthermore, the comparison of $V_{BD}$ for $\textit{type A}$
devices with different surface areas shows that $V_{BD}$ increases
slightly with decreasing the device area, possibly indicating that
the properties of SiO$_2$ in $\textit{type A}$ close to breakdown
are determined by small defects present in the film with rather
small probability. However, we cannot rule out that the differences
between the different sample populations originate from small
differences (2-3 nm) in the thickness of the SiO$_2$ layers. Note,
in fact, that the  leakage currents of these devices at low bias
have only small sample-to-sample fluctuations, suggesting that the
SiO$_2$ layers in $\textit{type A}$ devices are very
uniform.\\
To try to quantify better our observations and analyze the role
played by the specific fabrication technique and surface areas on
the device performance (e.g. breakdown field), we adopt a failure
analysis methodology \cite{Weibul1}. In what follows we provide a
statistical description of the breakdown probability introducing the
cumulative probability (P) as the probability of a device to
breakdown at a given voltage. From failure methodology, we notice
that possibly the most flexible distribution for the failure of a
population of samples, is the Weibull distribution
\cite{Weibul1,Weibul2} $P=1-exp[-(V_{BD}/V_{0})^{\beta}S/S_{0}]$
($S$ capacitor surface area, $S_{0}$ reference surface area, $\beta$
and $V_{0}$ are the Weibull parameters). The parameter $\beta$, also
known as Weibull shape parameter, determines the shape of the
probability density function -i.e. higher $\beta$ indicates
distributions with low dispersion of $V_{BD}$. $V_{0}$ is
the Weibull scale parameter, whose only effect is to scale the $V_{BD}$ distribution
(the larger  $V_{0}$, the more "stretched"  the
distribution). Depending on the value of the Weibull parameters,
this distribution mimics the behavior of other statistical
distributions such as the normal and the exponential. Given a sample population, the Weibull parameters provide a
quantitative measure of the failure probability. Both $\beta$ and $V_{0}$ are strongly
affected by the failure mechanism which can eventually be identified
when comparing the Weibull parameters for different sample
populations. For instance, a value of $\beta$ that does not depend on
the capacitor surface area -i.e., the variance does not change with the
surface area-, means that the microscopic mechanism of breakdown is common
to all the samples, independent of the specific area \cite{Weibul2}.\\
The good agreement between a fit to the Weibull distribution of the
cumulative probability for each different surface area and device
types shows that the data of each different device population is
well described by the Weibull distribution. To evaluate whether a
single Weibull scaling law can explain breakdown results for all the
different surfaces, we notice that -for \textit{type A}- we can fit
all the cumulative probability distributions with the same value
$\beta=52$, see Fig. \ref{Russo_fig4}b. This is made apparent by
plotting $ln(-ln(1-P))=ln(S/S_0)-\beta ln(V_0)+\beta ln(V_{BD})$ as
a function of $ln(V_{BD})$ for data sets corresponding to different
surface areas. Fig. \ref{Russo_fig4}b and c show clearly that all
the data are lined up on parallel linear slopes (i.e.,  $\beta$ is
the same in all these cases). This finding implies that  the
breakdown mechanism is the same for all studied surface areas. Note
that $V_{0}$ varies slightly with surface area, and that, as
mentioned above, we cannot exclude that the origin of these
variations is a small difference in the thickness of the SiO$_2$
layers for the different device populations
(a difference of 2-3 nm would suffice to explain this observation). \\
We notice that $\beta=52$ -estimated from the fit in Fig.
\ref{Russo_fig4}b and c- is comparable to values found for thermally
grown thin SiO$_2$ films, and it is compatible with failure of the
devices due to surface roughness \cite{SiO2roughness} probably being
transferred to the dielectric film from the Ti/Au substrate. This quantitative analysis make it possible to state
that electron-beam evaporated SiO$_2$, directly coated by a metallic layer without exposure to air has an essentially identical quality to that of thermally grown oxide. A
similar analysis of $P$ for $\textit{type B}$ devices, gives a
$\beta=3.3$ -i.e. a much higher dispersion of breakdown field (see Fig.
\ref{Russo_fig4}d). This small value for $\beta$ in $\textit{type
B}$ devices quantifies the much larger statistical
spread of the oxide properties in these devices. Since the only difference between $\textit{type A}$ and
$\textit{B}$ devices is the fabrication step of the SiO$_2$/Ti
interface, we conclude that exposure of the SiO$_2$ to air is indeed the cause for the poor insulating qualities.
Indeed it is well known that SiO$_2$ is an hygroscopic material, that easily absorbs humidity in air.
The humidity absorbed can affect the composition of the entire layer providing paths for
the leakage current, and creating weak spots at which breakdown occurs already at low voltage.\\

\section{Conclusions}

In conclusion we have briefly reviewed transport in double gated bilayer and trilayer
graphene devices.  Motivated by the need for large electric fields, we have conducted a statistical
study of the breakdown field for over 100 top gated structures
fabricated in different conditions and with different surface areas.
Adopting a failure analysis based on the Weibull distribution, we
show that the most reliable top gates are obtained when depositing
in SiO$_2$/Ti/Au without breaking the vacuum. Electron-beam SiO$_2$ layers evaporated
in these ways have insulating characteristics as good as those of thermally grown SiO$_2$ layers.

\subsection{Acknowledgments}
S. Russo acknowledges financial support from Stichting voor
Fundamenteel Onderzoek der Materie (FOM). M.F. Craciun acknowledges
financial support from the Japan Society for the Promotion of
Science, grant P07372. M. Yamamoto acknowledges Grant-in-Aid for
Young Scientists A (no. 20684011) and Exploratory Research for
Advanced Technology—Japan Science and Technology Agency
(080300000477). S. Tarucha acknowledges financial support from the
Grant-in-Aid for Scientific Research S (no. 19104007), B (no.
18340081) and Japan Science and Technology Agency—Core Research for
Evolutional Science and Technology. A.F. Morpurgo acknowledges
financial support from the Swiss National Science Foundation (grant
200021-121569) and from FOM.\\

\end{document}